\documentclass[runningheads]{llncs}
\usepackage[T1]{fontenc}
\usepackage{graphicx,verbatim}
\usepackage{graphicx}
\usepackage{booktabs}
\usepackage{multirow}
\usepackage{adjustbox}
\usepackage{xcolor}
\usepackage{amssymb}
\usepackage{amsmath}

\begin{document}
%
\title{DpDNet: An Dual-Prompt-Driven Network for Universal PET-CT Segmentation}
%

\author{Xinglong Liang\inst{1,2} \and
Jiaju Huang\inst{3} \and
Luyi Han\inst{1,2} \and
Tianyu Zhang\inst{1,2} \and
Xin Wang\inst{1,4} \and
Yuan Gao\inst{1,4} \and
Chunyao Lu\inst{1,2} \and
Lishan Cai\inst{1,4} \and
Tao Tan\inst{3,1}\thanks{Corresponding author}\and
Ritse Mann\inst{1,2}}
\authorrunning{X. Liang et al.}
\institute{
Department of Radiology, The Netherlands Cancer Institute, Plesmanlaan 121, 1066 CX, Amsterdam, The Netherlands
\and
Department of Radiology and Nuclear Medicine, Radboud University Medical Centre, Geert Grooteplein 10, 6525 GA, Nijmegen, The Netherlands
\and
Faculty of Applied Sciences, Macao Polytechnic University, 999078, Macao Special Administrative Region of China
\and
GROW School for Oncology and Developmental Biology, Maastricht University
Medical Centre, P. Debyelaan 25, 6202 AZ, Maastricht, The Netherlands
\email{taotan@mpu.edu.mo}\\}

\maketitle              
\begin{abstract}
PET-CT lesion segmentation is challenging due to noise sensitivity, small and variable lesion morphology, and interference from physiological high-metabolic signals. Current mainstream approaches follow the practice of one network solving the segmentation of multiple cancer lesions by treating all cancers as a single task. However, this overlooks the unique characteristics of different cancer types. Considering the specificity and similarity of different cancers in terms of metastatic patterns, organ preferences, and FDG uptake intensity, we propose DpDNet, a Dual-Prompt-Driven network that incorporates specific prompts to capture cancer-specific features and common prompts to retain shared knowledge. Additionally, to mitigate information forgetting caused by the early introduction of prompts, prompt-aware heads are employed after the decoder to adaptively handle multiple segmentation tasks. Experiments on a PET-CT dataset with four cancer types show that DpDNet outperforms state-of-the-art models. Finally, based on the segmentation results, we calculated MTV, TLG, and SUVmax for breast cancer survival analysis. The results suggest that DpDNet has the potential to serve as a valuable tool for personalized risk stratification, supporting clinicians in optimizing treatment strategies and improving outcomes.Code is available at \url{https://github.com/XinglongLiang08/DpDNet}.


\end{abstract}
\section{Introduction}
PET-CT (Positron emission tomography and computed tomography) is widely used for cancer diagnosis and prognosis, offering both anatomical and metabolic insights \cite{groheux2013performance,endo2006pet}. PET-CT typically covers whole-body imaging while the lesions of interest are usually small, and fluorodeoxyglucose (FDG) uptake is not confined to tumor tissues but also appears physiologically in areas like the urinary tract and myocardium \cite{haggstrom2024deep}, making PET-CT interpretation a time-consuming task for radiologists. Therefore, developing an accurate PET-CT-based automated lesion segmentation model to assist radiologists in diagnosis, evaluating patient treatment response, and assessing prognosis is valuable.

In recent years, deep learning-based models have significantly advanced medical image segmentation \cite{isensee2021nnu,huang2023stu,ye2023uniseg,hatamizadeh2021swin,hatamizadeh2022unetr,chen2024dynamic}. However, whole-body PET-CT segmentation still faces three major challenges. (1) Lesions in PET-CT range from small metastatic foci to large tumors, exhibiting diverse shapes, sizes, and locations, which makes automatic segmentation challenging. (2) The scarcity of PET-CT data leads to training on mixed PET-CT images without differentiating between cancer types, overlooking the specific characteristics of different cancers. (3) The lack of clinical evaluation of segmentation results, such as their impact on diagnostic accuracy and prognosis prediction, may prevent these models from gaining the trust of healthcare professionals and being adopted in practice.

Prompt engineering has the potential to be a promising tool for addressing these challenges in PET-CT segmentation. On one hand, it helps address the challenges posed by the small size and partially labeled nature of datasets \cite{liu2023clip}. On the other hand, prompt engineering integrates prior knowledge into models, enabling them to more effectively identify and segment regions of interest. A typical example is CLIP-driven universal models \cite{liu2023clip,zhang2023continual,zhang2024unimrisegnet}, which incorporate the text embeddings of all labels as external knowledge, obtained by feeding medical prompts into CLIP. However, CLIP, trained on natural language text, has limited capacity for understanding medical terminology, and even a slight change in wording can significantly impact its performance \cite{zhou2022learning}. Another groundbreaking work, DoDNet \cite{zhang2021dodnet,xie2023learning} can perform various segmentation tasks by utilizing task encoding and a controller to generate dynamic convolutions, but it suffers from delayed task awareness and the inability to leverage correlations among tasks. UniSeg \cite{ye2023uniseg,ye2024meduniseg} introduces a learnable universal prompt to capture correlations among tasks. This prompt is converted into a task-specific prompt and fed into the decoder, making the model task-aware early and improving task-specific performance. However, prematurely introducing prompts may lead to information forgetting, and using a shared segmentation head limits task-specific adaptability.

In this paper, we propose a dual-prompt-driven model for universal whole-body PET-CT segmentation. Previous methods overlook the heterogeneity among different cancer types, such as variations in metastatic patterns, metabolic activity as reflected by FDG uptake, and anatomical preferences for metastases. For instance, certain cancers like breast cancer often exhibit a distinct pattern of bone metastases \cite{kennecke2010metastatic}, while others, such as lung cancer, may preferentially spread to the brain or adrenal glands \cite{wood2014role}. These differences in biological behavior and imaging characteristics could impact the performance and generalizability of lesion segmentation models. On the other hand, metastases from different cancers are also interrelated, as they may share similar biological mechanisms of dissemination and exhibit overlapping imaging characteristics \cite{budczies2014landscape}. Based on these two considerations, we designed a dual-prompt strategy, comprising task-specific prompts and a common prompt. The specific prompts are tailored to capture the unique characteristics of metastases associated with each cancer type, while the common prompt leverages shared features and interrelations across different cancers to enhance the generalization of the model. These two prompts are fed into the decoder to introduce prompt information to the model early in the process. Additionally, we further designed prompt-aware heads to mitigate information forgetting and adaptively handle multiple segmentation tasks while avoiding a significant increase in model parameters. The head incorporates channel attention and multi-scale branches to enhance small-target feature extraction. Finally, we applied the segmentation results to breast cancer survival analysis to explore its potential clinical applications.


Our contributions are three-fold: (1) We design a dual-prompt strategy with specific and common prompts to capture cancer-specific metastatic patterns while leveraging shared features across different cancer types. (2) Prompt-aware heads are employed to replace the shared segmentation head, enabling adaptive task handling and enhancing small-target feature extraction via channel attention and multi-scale branches. (3) Predefined parameters are automatically calculated from segmentation results on a large dataset for breast cancer survival analysis, demonstrating our model's generalizability and its potential for risk stratification and survival prediction.

\begin{figure}
\includegraphics[width=\textwidth]{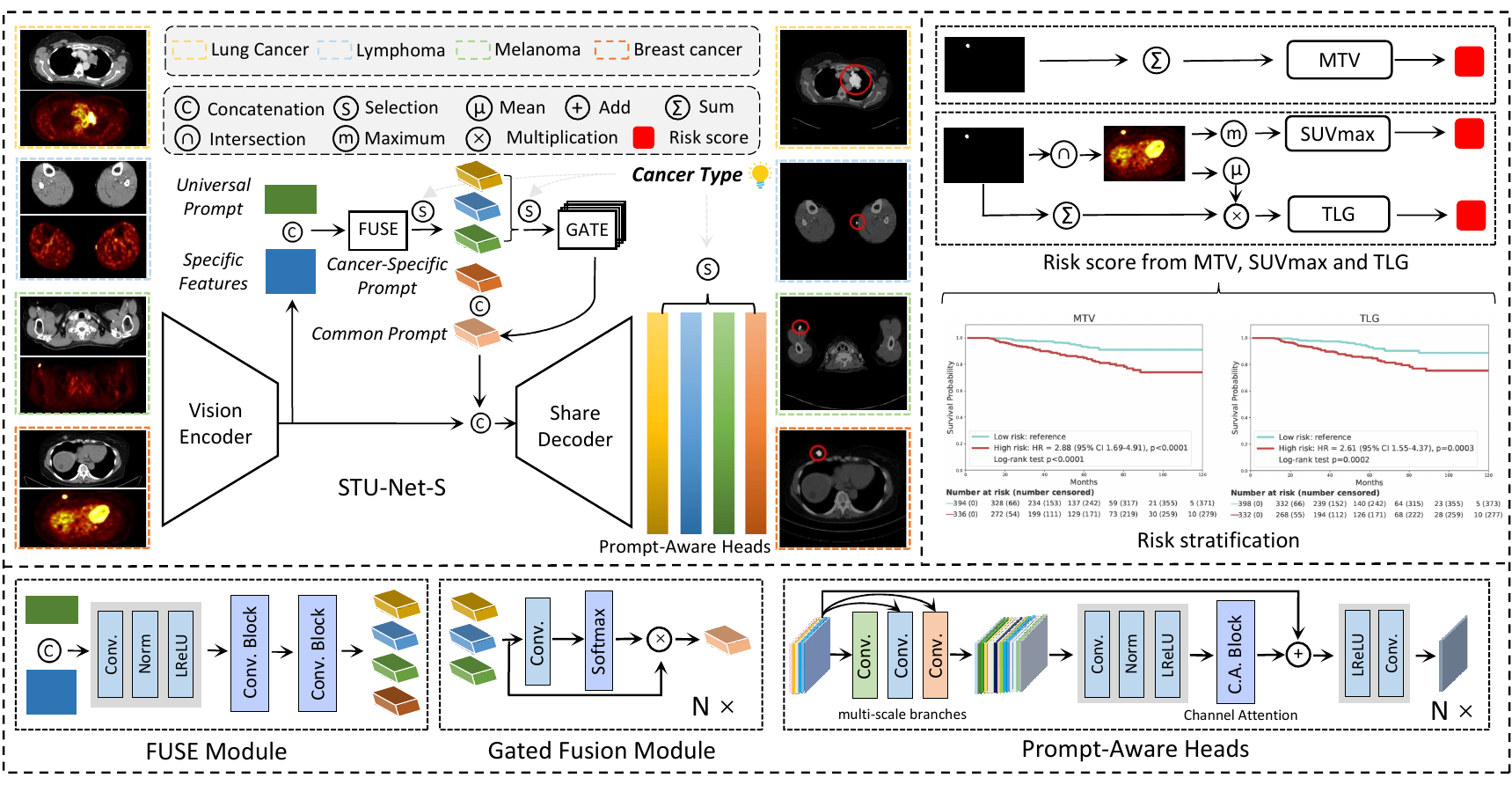}
\caption{Overview of the proposed DpDNet architecture for universal whole-body PET-CT segmentation. Cancer-specific information is captured using N (the number of cancer types) learnable prompts, which are processed via the FUSE Module to generate specific prompts. Simultaneously, common prompt is formed by N Gated Fusion Modules using fixed-order inputs from other cancer types. Both prompts are integrated into the shared decoder for segmentation. The architecture also incorporates prompt-aware heads, which are specialized segmentation heads tailored to each cancer type.} \label{fig1}
\end{figure}

\section{Methodology}
\subsection{Problem Definition}
Let \(\{C_1, C_2, \dots, C_N\}\) be \(N\) datasets of different cancer types. Here, \(C_i = \{X_{ij}, Y_{ij}\}_{j=1}^{n_i}\) represents that the \(i\)-th dataset has a total of \(n_i\) image-label pairs, and \(X_{ij}\) and \(Y_{ij}\) are the image and the corresponding ground truth, respectively. Typically, PET-CT-based cancer segmentation models are trained either by treating all cancer datasets as a single task or by focusing on a specific cancer type. The former approach fails to capture the distinct characteristics of different cancers, while the latter is constrained by limited data availability. To address these challenges, we propose a universal PET-CT segmentation framework, termed DpDNet. The overall architecture of DpDNet is shown in Fig. \ref{fig1}, with details introduced in the following subsections.

\subsection{Encoder-decoder backbone}
The backbone of DpDNet is based on STU-Net \cite{huang2023stu}, a scalable and transferable large-scale medical segmentation model. STU-Net is based on the nnU-Net architecture but incorporates key modifications to the encoder, decoder, and upsampling processes. Residual connections are added to the basic blocks to address gradient diffusion issues, while downsampling is integrated into the first residual block of each stage using a dual-branch design for a more streamlined architecture. Additionally, for upsampling, transpose convolution is replaced with nearest neighbor interpolation followed by a 1$\times$1$\times$1 convolution, enhancing efficiency and eliminating weight mismatches across tasks. Considering the need for efficiency and reduced computational cost, we adopt the smallest variant, STU-Net-S, as the backbone network, which contains only 14 million parameters. 
\subsection{Cancer-specific prompt and common prompt}
As mentioned earlier, different PET-CT cancer segmentation tasks are highly correlated. While models like UniSeg and CLIP-Driven models account for these correlations, their non-decoupled approach may weaken task-specific features or fail to fully utilize shared ones. In contrast, our method employs cancer-specific and common prompts to explicitly separate cancer-specific and shared features, capturing correlations while preserving their uniqueness. Following the method used in UniSeg, we define the shape of the universal prompt as \(F_{\text{uni}} \in \mathbb{R}^{N \times \frac{D}{16} \times \frac{H}{32} \times \frac{W}{32}}\), where \(N\) is the number of tasks.

\begin{equation}
\{F_{\text{can}1}, \dots, F_{\text{can}i}, \dots, F_{\text{can}N}\} = \text{Split}\left(f_1\left(\text{cat}(F_{\text{uni}}, F)\right)\right)^N
\label{eq:common_prompts}
\end{equation}

Where \(F_{\text{can}i}\) denotes the prompt features belonging to the \(i\)-th cancer and \(F\) denotes the sample-specific features from the bottleneck. \(\text{cat}(\cdot, \cdot)\) represents a concatenation operation, \(f(\cdot)\) denotes the feed-forward processes, and \(\text{Split}(\cdot)^N\) indicates splitting features along the channel to obtain \(N\) cancer-specific prompts. Then, we select the target prompt according to the ongoing task from \(\{F_{\text{can}1}, \dots, F_{\text{can}N}\}\). Subsequently, we perform gated fusion on the remaining prompts, excluding the current task, to obtain \(F_{\text{com}i}\), defined as:

\begin{equation}
F_{\text{ini}} = f_{2i}(\text{cat}(F_{\text{can}1}, \dots, F_{\text{can}i-1}, F_{\text{can}i+1}, \dots, F_{\text{can}N}))
\label{eq:input_features}
\end{equation}
\begin{equation}
F_{\text{com}i} = \text{Softmax}(F_{\text{cat}(F_{\text{can}1}, \dots, F_{\text{can}i-1}, F_{\text{can}i+1}, \dots, F_{\text{can}N})}) \odot F_{\text{ini}}
\label{eq:common_features}
\end{equation}

Finally, we concatenate \(F\), the selected \(F_{\text{cani}}\) and common prompt \(F_{\text{comi}}\) as the decoder input. In this way, cancer-related prior information is introduced into the model to enhance the training of the entire decoder.

\subsection{Prompt-aware Heads}

Unlike the dynamic convolution heads in DoDNet and CLIP-Driven models, we designed prompt-aware heads, which replace the final stage of the decoder by selecting segmentation heads based on cancer-type prompts. Prompt-aware heads incorporate channel attention to capture inter-channel relationships, multi-scale branches improve feature extraction: the 1$\times$1$\times$1 branch preserves original information to prevent loss, the 3$\times$3$\times$3 branch captures local features for small-target edges and textures, and the 5$\times$5$\times$5 convolution branch expands the receptive field. Compared to dynamic convolution heads (which apply only a single convolution after the decoder and may not fully capture diverse cancer characteristics), our design adapts earlier to different cancer types, enabling better feature learning while also simplifying training, especially for limited datasets. In contrast to methods that assign a separate decoder for each task, our approach balances task adaptability and efficiency without significantly increasing model parameters.
\section{Experiments and Results}
\subsection{Datasets and Evaluation Metric}
\subsubsection{Datasets} For this study, we collected whole-body PET-CT data from four types of cancer: lung cancer, lymphoma, breast cancer, and melanoma to train and test DpDNet. The lung cancer, lymphoma, and melanoma datasets were sourced from AutoPET \cite{gatidis2022whole}, while the breast cancer dataset came from our private collection. The combined dataset includes a total of 697 cases, with 557 cases used for training and 140 cases for validation. For the survival prediction, we collected 1,210 nonmetastatic breast cancer patients, without ground truth. Median values of SUVmax, MTV, and TLG were computed from 480 patients and used to stratify the remaining 780 patients into high- and low-risk groups. 
\subsubsection{Evaluation Metric} We evaluated the segmentation performance using four metrics: Dice similarity coefficient (DSC) and Intersection over Union (IoU). For the survival prediction, we used C-index and Hazard Ratio (HR) as the evaluation metric.
\subsection{Implementation Details}
During training, we adopted the SGD optimizer and set the batch size to 2, the initial learning rate to 0.0001, the default patch size to \(112 \times 160 \times 128\) (while SwinUNet used \(128 \times 128 \times 128\) due to network limitations), and the maximum training epoch to 1000 with a total of 250,000 iterations. All experiments were conducted on the NVIDIA Quadro RTX A6000 GPU.
\subsection{Results}
\subsubsection{Comparing to Non-Prompt-Based and Prompt-Based Models} Our DpD-Net was compared with the state-of-the-art Non-Prompt-Based and Prompt-Based models. Additionally, we trained nnU-Net separately on each cancer type for comparison. The Non-Prompt-Based models includes nnU-Net \cite{isensee2021nnu}, STU-NET \cite{huang2023stu}, UNETR \cite{tang2022self}, 3DUX-Net \cite{lee20223d}, SwinUNETRV2 \cite{tang2022self}, U-Mamba \cite{ma2024u}. The Prompt-Based models includes CLIP-driven Universal Model \cite{liu2023clip}, DoDNet\cite{zhang2021dodnet}, UniSeg \cite{ye2023uniseg}. To balance network complexity and performance, we adopted STU-Net-S as the backbone for all prompt-based models. As shown in Table \ref{result}, our method achieves the highest average segmentation performance across multiple cancer types. Specifically, the average DSC surpasses the second-best method by 1.32\% and the average IoU improves by 1.56\%, while our approach achieves the best segmentation results in lung cancer, lymphoma, and breast cancer. Additionally, our model maintains a parameter size of approximately 15M and FLOPs around 140G. These results highlight the advantages of our method in both segmentation accuracy and efficiency. Moreover, prompt-based models generally outperform non-prompt-based models, demonstrating the effectiveness of the cancer-level prompts we introduced for PET-CT segmentation tasks across whole-body lesions. 
\subsubsection{Learnable Prompt Visualization} 
\begin{figure}
\centering
\includegraphics[width=1\textwidth]{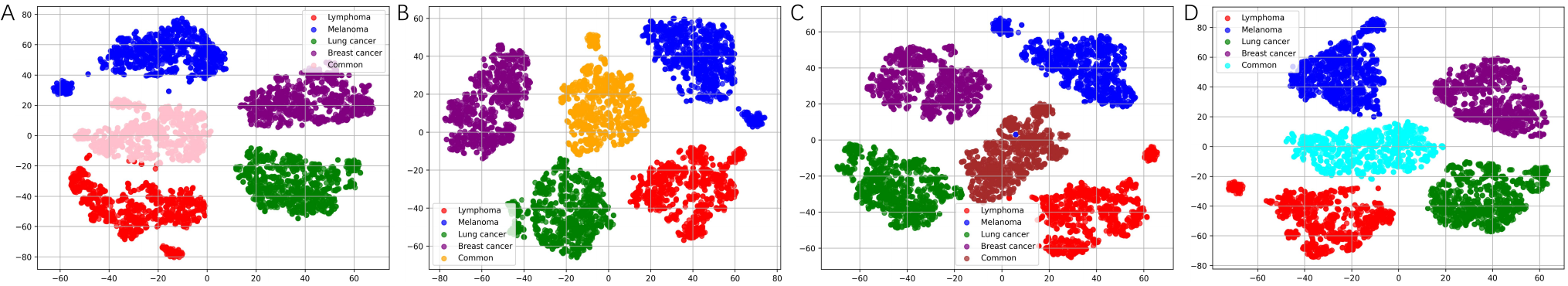}
\caption{T-SNE visualization of learnable cancer-specific and common prompts for (A) Lung Cancer, (B) Lymphoma, (C) Melanoma, and (D) Breast Cancer. The common prompt for each cancer type is derived from the fusion of cancer-specific prompts from other cancers, leading to distinct representations. To illustrate these differences, we present four separate visualizations.} \label{fig2}
\end{figure}

We conducted a visualization analysis of the learnable cancer-specific and common prompts using T-SNE. As shown in Fig. \ref{fig2}, red, blue, purple, and green points represent different cancer-specific prompts, with the common prompts positioned centrally among them, indicating that they encode shared features across cancer types. These findings demonstrate that our method effectively separates cancer-specific features, capturing their correlations while preserving their distinctiveness.

\begin{table}[h!]
\centering
\caption{Comparison of different methods for cancer lesion segmentation with Weighted Averages. Bold values indicate the highest score, and underlined values indicate the second highest. All comparison results yielded $p<0.05$. ("Single" denotes single-task training, "+" indicates nnU-Net with residual connections, and "*" represents STU-Net-B).}
\label{result}
\begin{adjustbox}{max width=\textwidth}
\begin{tabular}{llccccccccccccc}
\toprule
\multirow{3}{*}{\textbf{Category}} & \multirow{3}{*}{\textbf{Metrics}} & \multicolumn{8}{c}{\textbf{Non-Prompt-Based Methods}} & \multicolumn{5}{c}{\textbf{Prompt-Based Methods}} \\ 
\cmidrule(lr){3-11} \cmidrule(lr){12-15}
 &  & \textbf{Single} & \textbf{nnU-Net} & \textbf{nnU-Net\textsuperscript{+}} & \textbf{STU-Net} & \textbf{STU-Net*} & \textbf{UNETR} & \textbf{SwinUNET} & \textbf{U-Mamba} & \textbf{3DUX-Net} & \textbf{Dod-Net} & \textbf{CLIP-driven} & \textbf{UniSeg} & \textbf{Proposed} \\
\midrule
\multirow{2}{*}{\textbf{Lung cancer}} 
& DSC       & 80.45 & 78.57 & 78.83 & 79.55 & \underline{80.68} & 72.92 & 75.94 & 79.76 & 80.05 & 79.26 & 80.39 & 80.45 & \textbf{80.81} \\
& IoU       & 68.31 & 65.86 & 66.07 & 67.20 & \underline{68.51} & 59.37 & 62.99 & 67.67 & 68.05 & 66.91 & 68.49 & 68.34 & \textbf{68.98} \\
\midrule
\multirow{2}{*}{\textbf{Lymphoma}} 
& DSC       & 71.30 & 73.66 & 71.93 & 70.15 & 74.07 & 68.85 & 70.65 & 72.49 & 73.01 & \underline{76.80} & 74.38 & 75.87 & \textbf{77.76} \\
& IoU       & 59.57 & 61.51 & 59.75 & 58.15 & 61.90 & 56.92 & 58.83 & 60.18 & 61.08 & \underline{64.92} & 62.95 & 64.40 & \textbf{66.38} \\
\midrule
\multirow{2}{*}{\textbf{Melanoma}} 
& DSC       & 72.19 & 72.44 & 75.46 & 70.53 & 75.03 & 61.23 & 68.71 & 70.34 & \textbf{76.96} & 75.57 & 75.82 & 72.70 & \underline{76.44} \\
& IoU       & 60.43 & 60.17 & 63.70 & 58.32 & 62.52 & 48.27 & 55.87 & 58.38 & \underline{64.93} & 64.10 & 64.36 & 61.09 & \textbf{65.15} \\
\midrule
\multirow{2}{*}{\textbf{Breast cancer}} 
& DSC       & \underline{65.89} & 60.23 & 61.51 & 62.65 & 63.26 & 57.02 & 61.79 & 61.60 & 64.21 & 64.41 & 64.45 & 64.85 & \textbf{66.27} \\
& IoU       & \underline{51.76} & 45.45 & 47.14 & 48.17 & 48.91 & 42.78 & 47.68 & 47.36 & 49.61 & 50.09 & 50.10 & 50.58 & \textbf{51.85} \\
\midrule
\multirow{2}{*}{\textbf{AVG}} 
& DSC       & 72.07 & 70.66 & 71.54 & 70.25 & 72.79 & 64.33 & 68.81 & 70.46 & 73.22 & \underline{73.55} & 73.35 & 72.93 & \textbf{74.87} \\
& IoU       & 59.58 & 57.63 & 58.74 & 57.42 & 59.92 & 51.11 & 55.82 & 57.76 & 60.50 & \underline{61.00} & 60.99 & 60.50 & \textbf{62.56} \\
\midrule
\textbf{Parameters} 
& M         & 30.79 & 30.79  & 46.71  & 14.55  & 58.16  & 96.19  & 46.49  & 42.12  & 83.81  & 15.14  & 15.34  & 15.12  & 15.19 \\
\textbf{FLOPs} 
& GFLOPs    & 526.29 & 526.29 & 833.94 & 138.70 & 548.61 & 196.92 & 202.58 & 1067.69 & 1509.02 & 139.26 & 139.26 & 138.79 & 138.79 \\
\bottomrule
\end{tabular}
\end{adjustbox}
\end{table}

\subsubsection{Ablation Study} Table \ref{tab:ablation_study} shows the contribution of each component to the model performance. Compared to the baseline (70.24\%), adding the cancer-specific prompt increases DSC to 72.93\%, while further introducing the common prompt boosts it to 73.79\%. Prompt-aware heads (PA-Heads) alone achieve 73.92\%, highlighting their individual effectiveness. Combining all components yields the highest DSC of 74.87\%, emphasizing their complementary roles.


\begin{table}[ht]
\centering
\caption{Ablation study results on the average DSC (\%).}
\label{tab:ablation_study}
\resizebox{0.8\textwidth}{!}{ 
\begin{tabular}{cccc|ccccc}
\toprule
\multicolumn{4}{c}{\textbf{Experimental Setting}} & \multicolumn{5}{c}{\textbf{Category}} \\
\cmidrule(r){1-4} \cmidrule(l){5-9}
Baseline & S-Prompt & C-Prompt & PA-Heads & Lung & Lymphoma & Melanoma & Breast & AVG \\
\midrule
\checkmark  & & & & 79.55 & 70.15 & 70.53 & 62.65 & 70.24 \\
\checkmark  & \checkmark & & & 80.45 & 75.87 & 72.70 & 64.85 & 72.93 \\ 
\checkmark  & \checkmark& \checkmark & & \textbf{81.05} & 76.63 & 74.59 & 64.98 & 73.79 \\
\checkmark  &  &  & \checkmark & 79.66 & 76.48 & 75.54 & 65.71 & 73.92 \\
\checkmark  & \checkmark & \checkmark & \checkmark & 80.81 & \textbf{77.76} & \textbf{76.44} & \textbf{66.27} & \textbf{74.87} \\
\bottomrule
\end{tabular}
}
\end{table}

\subsubsection{Survival analysis}
Biomarkers derived from radiological imaging have been effectively utilized to predict breast cancer prognosis \cite{gao2024explainable,zhang2023radiomics}. Several studies have investigated the prognostic value of metabolic activity (SUVmax), tumor volume (MTV), and total lesion glycolysis (TLG) \cite{yue2015stratifying,il2021prognostic,bouron2022prognostic,quartuccio202318f}. In this study, we leverage our proposed method to perform inference on a large dataset without ground truth and automatically extract predefined parameters for survival analysis.
\begin{table}[htbp]
\centering
\caption{Comparison of three best-performing segmentation models for Survival Prediction (C-index \%)}
\label{tab:comparison_sur}
\renewcommand{\arraystretch}{1.2} 
\begin{tabular}{lccc}
\toprule
\textbf{Method} & \textbf{MTV} & \textbf{TLG} & \textbf{SUVmax} \\
\midrule
Dod-Net    &  68.51 [61.79, 75.30]& 66.12 [59.12, 72.35]& 62.58 [56.09, 68.44] \\
CLIP-driven  & 68.83 [61.62, 75.69] & 66.07 [59.23, 72.43] & 62.44 [56.10, 68.97] \\
Proposed     & \textbf{69.33} [62.05, 75.83] & \textbf{66.63} [59.52, 72.84]& \textbf{62.74} [55.65, 68.89] \\
\bottomrule
\end{tabular}
\end{table}


\begin{figure}\centering
\includegraphics[width=\textwidth]{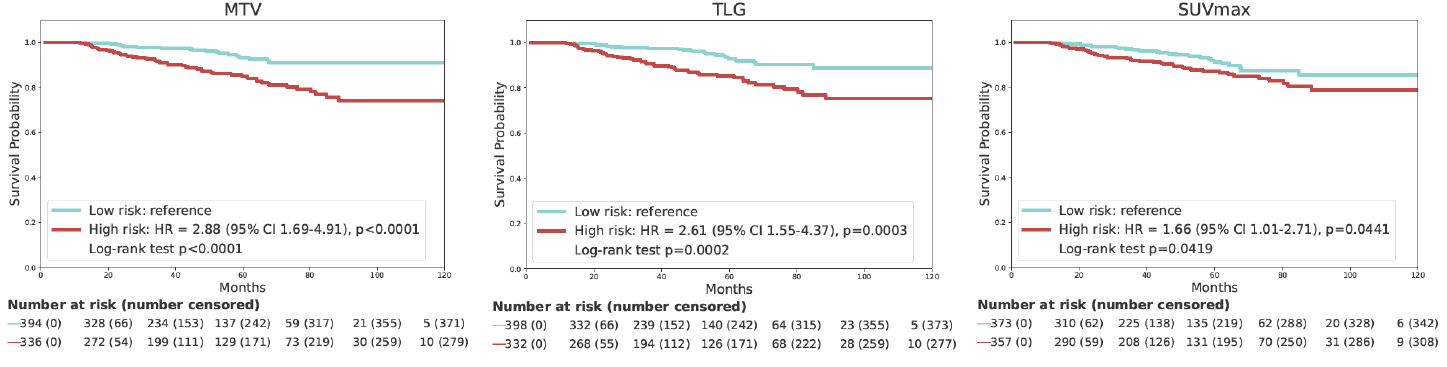}
\caption{Kaplan-Meier survival curves stratified by MTV, TLG, and SUVmax, showing the survival differences between high-risk and low-risk groups.} \label{fig3}
\end{figure}

The results are presented in Table \ref{tab:comparison_sur}. We selected the three best-performing segmentation models to calculate MTV, TLG, and SUVmax and evaluated their prognostic performance using the C-index. The findings indicate that better segmentation accuracy enhances survival prediction precision, demonstrating the generalizability of our model. Patients were then stratified into high-risk and low-risk groups based on the median values of SUVmax, MTV, and TLG. As shown in Fig. \ref{fig3}, patients in the high-risk group had worse outcomes than those in the low-risk group. A Cox proportional hazards model was applied based on this stratification, revealing that MTV provided the strongest prognostic stratification. Specifically, for overall survival, MTV had the highest hazard ratio (HR = 2.88), compared to TLG (HR = 2.61) and SUVmax (HR = 1.66).

Our study demonstrates the potential of the proposed universal segmentation model as a reliable and efficient clinical tool for breast cancer prognosis. By automatically extracting key parameters from PET-CT, it eliminates the need for time-consuming manual annotations. The model stratifies patients into high- and low-risk groups, enabling targeted monitoring and early interventions while minimizing unnecessary treatments. These findings underscore its value in enhancing cancer diagnosis, treatment planning, and patient management.

\section{Conclusion}
In this study, we propose a dual-prompt-driven model designed for whole-body PET-CT segmentation across multiple cancer types. To address the heterogeneity and interrelations among different cancers, we introduce a dual-prompt strategy that integrates cancer-specific and common prompts, enabling the model to distinguish unique metastatic patterns while leveraging shared features. Additionally, we incorporate prompt-aware heads to adaptively handle multiple segmentation tasks while maintaining efficiency. With these designs, DpDNet achieves state-of-the-art performance across multiple PET-CT datasets and outperforms various existing segmentation models. Furthermore, its application to breast cancer survival analysis demonstrates its potential for risk stratification and clinical decision support. These findings indicate that DpDNet could be a scalable and effective tool for advancing personalized cancer management.

\section*{Acknowledgements}
Xinglong Liang was funded by Chinese Scholarship Council (CSC) scholarship. This work was supported by the Science and Technology Development Fund of Macao (0041/2023/RIB2).

\section*{Disclosure of Interests}
The authors declare no competing interests.
%
%
%
\bibliographystyle{splncs04}
\bibliography{mybibliography}
\end{document}